\documentclass{article}

\usepackage{microtype}
\usepackage{graphicx}
\usepackage{subfigure}
\usepackage{booktabs} % for professional tables

\usepackage{hyperref}

\usepackage[accepted]{icml2025}

\usepackage{amsmath}
\usepackage{amssymb}
\usepackage{mathtools}
\usepackage{amsthm}
\usepackage{bm}
\usepackage{colortbl}
\usepackage{caption}
\usepackage{bm}

\usepackage[capitalize,noabbrev]{cleveref}

\theoremstyle{plain}

\theoremstyle{definition}

\theoremstyle{remark}

\newcommand{\defeq}{\coloneqq}

\newcommand{\E}{\mathbb{E}}

\newcommand{\Eb}[2]{\E_{#1}\!\left[#2\right]}

\newcommand{\bI}{\mathbf{I}}

\newcommand{\bzero}{\mathbf{0}}

\newcommand{\bx}{\mathbf{x}}

\newcommand{\bepsilon}{{\boldsymbol{\epsilon}}}

\usepackage[textsize=tiny]{todonotes}

\definecolor{Gray}{gray}{0.9}

\icmltitlerunning{}

\begin{document}

\twocolumn[
\icmltitle{Bootstrap-GS: Self-Supervised Augmentation for High-Fidelity \\ Gaussian Splatting }



\icmlsetsymbol{equal}{*}
 %Kerui Ren, Jie Ou, Lei Wang, Jiaji Wu, Jun Cheng
\begin{icmlauthorlist}
\icmlauthor{Yifei Gao}{equal}
\icmlauthor{Kerui Ren}{equal}
\icmlauthor{Ou Jie}{equal}
\icmlauthor{Lei Wang$\dag$}{}
\icmlauthor{Jiaji Wu}{}
\icmlauthor{Jun Cheng}{}

\end{icmlauthorlist}

%\icmlaffiliation{yyy}{Department of XXX, University of YYY, Location, Country}
%\icmlaffiliation{comp}{Company Name, Location, Country}
%\icmlaffiliation{sch}{School of ZZZ, Institute of WWW, Location, Country}

%\icmlcorrespondingauthor{Firstname1 Lastname1}{first1.last1@xxx.edu}
%\icmlcorrespondingauthor{Firstname2 Lastname2}{first2.last2@www.uk}

\icmlkeywords{Machine Learning, ICML}

\vskip 0.3in

\renewcommand\twocolumn[1][]{#1}%
% \maketitle
\vspace{-6mm}%
\begin{center}
    \centering
	\includegraphics[width=1.\textwidth]{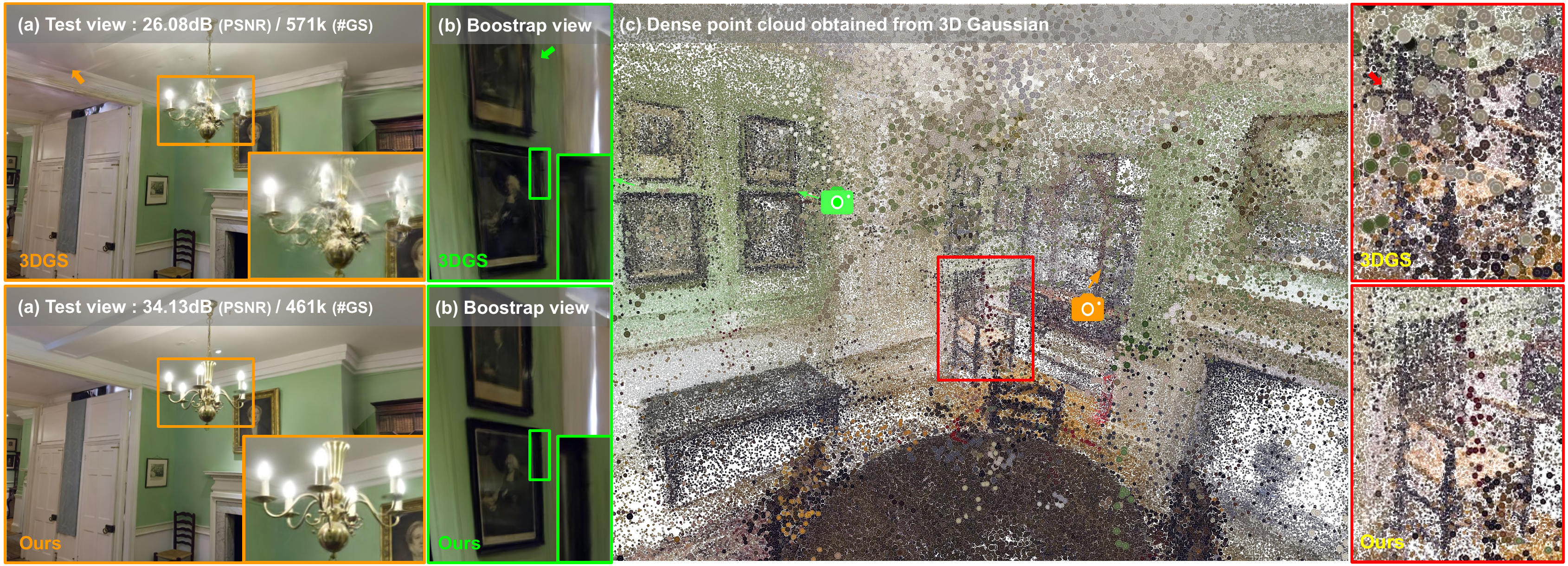}
	\vspace{-6mm}
	\captionof{figure}{By addressing the common issue of training sampling deficiency in 3D reconstruction, our bootstrap technique significantly reduces artifacts in novel-view renderings and enables 3D-GS to render superior results with clear and more structured point clouds.}
    \label{fig:teaser}
    \vspace{2mm}%
\end{center}%

]




\begin{abstract}
Recent advancements in 3D Gaussian Splatting (3D-GS) have established new benchmarks for rendering quality and efficiency in 3D reconstruction. However, 3D-GS faces critical limitations when generating novel views that significantly deviate from those encountered during training. Moreover, issues such as dilation and aliasing arise during zoom operations. These challenges stem from a fundamental issue: training sampling deficiency. In this paper, we introduce a bootstrapping framework to address this problem. Our approach synthesizes pseudo-ground truth from novel views that align with the limited training set and reintegrates these synthesized views into the training pipeline. Experimental results demonstrate that our bootstrapping technique not only reduces artifacts but also improves quantitative metrics. Furthermore, our technique is highly adaptable, allowing various Gaussian-based method to benefit from its integration.
\end{abstract}
\section{Introduction}
\label{sec:intro}
Lately, 3D Gaussian Splatting (3D-GS)~\cite{kerbl20233dgs} has emerged as a cutting-edge method in rendering, demonstrating unparalleled quality and efficiency. This approach has attracted considerable attention, enhancing a variety of applications such as VR interactions~\cite{xie2023physgaussian}, drivable human avatars~\cite{qian2023gaussianavatars}, and navigation through large-scale urban scenes~\cite{zhou2023drivinggaussian}. Moreover, its utility has expanded to visual effects such as splashing~\cite{feng2024gaussiansplashing}, style transformation~\cite{liu2024stylegs}, and object segmentation and editing~\cite{ye2024gsgrouping}, indicating significant commercial potential.

Nonetheless, due to the inherent characteristics of 3D-GS’s rendering process, artifacts such as distortion, alias, and high-frequency outliers continue to emerge under novel viewpoints. To address these issues, previous approaches have often focused on improving model architectures and rendering processes. Specifically, Mip-Splatting~\cite{yu2023mipsplat} uses filters to eliminate Gaussian primitives that could cause artifacts and to prevent aliasing. GaussianPro~\cite{cheng2024gaussianpro} standardizes the Gaussian normals to achieve a more uniform and smoother distribution. Scaffold-GS~\cite{lu2023scaffold} replaces Spherical Harmonics for color representation with multi-layer perception (MLP) prediction and introduces offsets to represent nearby Gaussian primitives. Additionally, Octree-GS~\cite{ren2024octreegs} incorporates the concept of Level of Detail, not only refining the structure of Gaussians but also reducing their volume. 

Despite their achievements, they all overlook a fundamental issue: \textbf{training sampling deficiency}. Specifically, 3D-GS relies on matching scenes with Gaussian distributions; however, because the training data only provides incomplete scenes, the resulting reconstructions exhibit prominent artifacts as shown in Figure~\ref{fig:teaser}, which cannot be remedied by merely modifying the model architecture or its rendering process. Moreover, obtaining additional data—whether authentic or just “largely aligned”~\cite{chen2024text}—is often infeasible in open scenes~\cite{barron2022mipnerf360}. These constraints prompt the central question of this paper: \textbf{can we leverage the reconstructed scenes themselves to extend into those unknown, unexplored viewpoints?}

In this paper, we propose a bootstrapping framework that enables the creation of new perspective data with high consistency to the current scene. By leveraging the partially reconstructed scene, we perform bootstrapping by adjusting the camera angles to obtain multiple renderings from new viewpoints. These renderings are then processed through a diffusion model, followed by multi-view and multi-sampling averaged loss optimization.
%This approach significantly enhances the consistency of multi-view generation while preserving the original scene information. 
Through extensive experiments, we demonstrate that our method not only significantly improves various metrics but also strengthens existing details and generates consistent new scene information, even from viewpoints that are far divergent from the training datasets. Additionally, it effectively reduces artifacts such as distortion and aliasing. Furthermore, the underlying concept of our method is universally applicable and plug-and-play, allowing it to enhance a range of previously proposed architectures. And it is not limited to the 3D-GS framework.

In summary, the main contributions of our method are: 
1) We propose an innovative self-supervised augmentation strategy for high-fidelity novel view synthesis.
2) We demonstrate the plug-and-play capability of our method and its extensive applicability.
3) Our approach achieves comprehensive advancements over prior state-of-the-art efforts and the vanilla 3D-GS.
\section{Related Work}

% \subsection{Multi-view 3D Reconstrcution}
\subsection{Novel View Synthesis}
Early methods such as NeRF~\cite{martin2021nerf} typically employ an MLP to serve as a global approximator for 3D scene geometry and appearance. These approaches~\cite{barron2022mipnerf360,barron2023zip} directly feed spatial coordinates (along with the viewing direction) into the MLP to predict point-wise attributes. Although they can produce high-quality renderings, they require excessively long computation times. Grid-based techniques, including interpolation~\cite{fridovich2022plenoxels,muller2022instant} and tensor factorization~\cite{chen2022tensorf}, have also been explored. However, aside from their speed advantage, these methods still struggle to effectively represent empty space and handle divergent viewpoints. Recently, the point-based method 3D-GS was introduced to achieve real-time rendering with high-quality results and fine-scale detail. It models the scene using 3D Gaussians, which are optimized in a volumetric manner and then projected to 2D.

%\vspace{-0.2cm}
\subsection{Diffusion-based Sparse View Reconstrcution}
Starting from text-to-3D reconstruction, recent efforts have increasingly leveraged high-quality 2D diffusion models~\cite{nichol2021improved} for 3D tasks. Pioneering works such as DreamFusion~\cite{poole2022dreamfusion} introduced a distillation process that transforms a 2D text-to-image generation model into a 3D generator guided by textual prompts, embedding geometry and view information within the prompts. This approach has inspired a wave of subsequent studies~\cite{chen2024text, tang2023dreamgaussian,liu20243dgs-enhancer}. However, these methods are restricted to object reconstruction and can only operate on pre-trained objects.  

For real-world sparse-view reconstruction using diffusion models, GaussianObject~\cite{yang2024gaussianobject} employs diffusion models solely for constructing coarse 3D-GS representations of objects, relying on a separate refinement model for fine-detail consistency. Similarly, StreetGS~\cite{yu2024sgd} integrates diffusion models with multimodal data to regulate point clouds of 3D-GS. Although it operates on full scenes, it still avoids directly addressing multi-view consistency at a fine-detail level.  

In this paper, we propose a novel method that directly tackles this challenge, enabling consistent multi-view reconstruction with diffusion models.

%The fundamental challenge lies in the limited availability of open-scene datasets and their unique and complex scene characteristics. These factors prevent diffusion models from adequately learning the scene structure and generating multi-view images with consistent quality. 

\section{Preliminaries}
\paragraph{3D Gaussian Splatting}
3D-GS~\cite{kerbl20233dgs} models the scene using a collection of anisotropic 3D Gaussians which are further rendered to images using the splatting-based rasterization technique. For each 3D Gaussian $G$, it is defined as:
\begin{equation}
\label{eq:gaussian}
G(\mathbf{x})=e^{-\frac{1}{2}(\mathbf{x}-\bm{\mu})^T \bm{\Sigma}^{-1}(\mathbf{x}-\bm{\mu})},
\end{equation}
In this context, $x$ represents any arbitrary position within the 3D scene, and $\Sigma$ signifies the covariance matrix of the 3D Gaussian. The covariance matrix $\Sigma$ is constructed utilizing a scaling matrix $S$ and a rotation matrix $R$, ensuring that it remains positive semi-definite: $\Sigma = RSS^TR^T$. To render an image from a given viewpoint, the color of each pixel $\mathbf{p}$ is calculated by blending $N$ ordered Gaussians $\left\{G_i \mid i=1, \cdots ,N\right\}$ overlapping $\mathbf{p}$ with learned opacity and color for each Gaussian.

\paragraph{Diffusion Model}
Diffusion models~\citep{sohl2015deep} are latent variable models of the form $p_\theta(\bx_0) \defeq \int p_\theta(\bx_{0:T}) \,d\bx_{1:T}$, where $\bx_1, \dotsc, \bx_T$ are latent variables with the same dimensionality as the data $\bx_0\sim q(\bx_0)$. 
The \emph{forward process}, synonymous with the \emph{diffusion process}, is formulated as a Markov chain, which incrementally incorporates Gaussian noise into the data according to a pre-specified variance schedule. Conversely, the \emph{reverse process} corresponds to the joint distribution $p_\theta(\bx_{0:T})$ and is also defined as a Markov chain with Gaussian transitions that are learned from data, beginning with an initial distribution  $p(\bx_T)=\mathcal{N}(\bx_T; \bzero, \bI)$. In this process, the original distribution at time step $0$ is gradually recovered from time step $T$. And the training is performed by optimizing the usual variational bound on negative log likelihood. If all the conditionals are modeled as Gaussians with trainable mean functions and fixed variances, the training objective can be simplified to:
\begin{footnotesize}
\begin{align}
 L_\mathrm{simple}(\theta) \defeq \Eb{t, \bx_0, \bepsilon}{ \left\| \bepsilon - \bepsilon_\theta(\sqrt{\bar\alpha_t} \bx_0 + \sqrt{1-\bar\alpha_t}\bepsilon, t) \right\|^2} \label{eq:dm_obj}
\end{align}
\end{footnotesize}
where $\epsilon_\theta$ is a learned noise prediction function, $\bar\alpha_t$ is the schedule factor, and $\bepsilon \sim \mathcal{N}(0, I)$ is a normal noise. For more details, please refer to references~\citep{ho2020denoising,song2020denoising}.
\begin{figure*}[!t]
    \centering
     \includegraphics[width=1\linewidth]{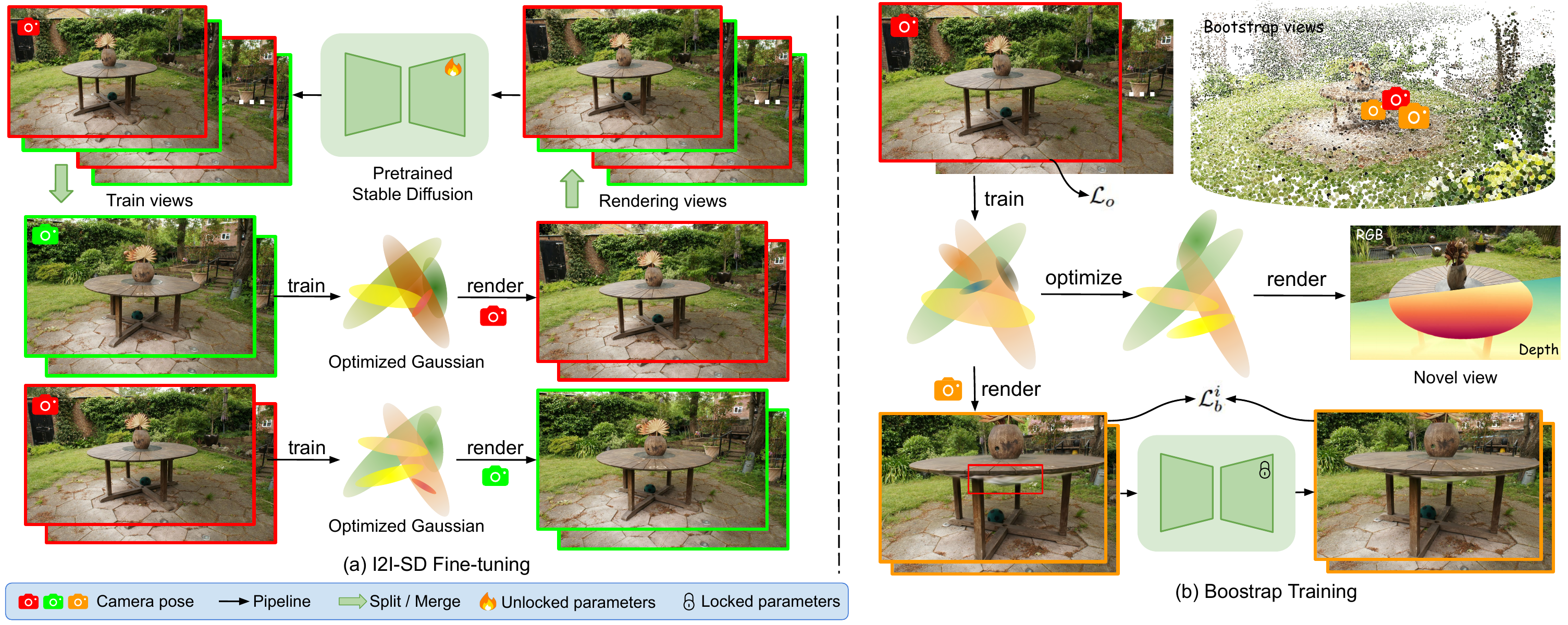}
    \caption{\textbf{Overall pipeline}. (a) To simulate the artifacts present in novel-view renderings, we first train a 3D-GS model using only half of the training set with limited time steps and then render the remaining half. This process is then repeated for the other half of the dataset. Finally, we obtain the fine-tuning data for diffusion models, where the renderings serve as \( x_t \) and the ground truth as \( x_0 \).  (b) For each training camera, we bootstrap several novel-view cameras and then acquire their corresponding renderings. After diffusion regeneration, these renderings are reintegrated into the training process, where multiple bootstrap cameras, along with a single training camera, are used to compute the bootstrapping loss for each training iteration.}
    \label{fig:pipeline}
    \vspace{-1pt}
\end{figure*}

\section{Method}

In this section, we first define the ill-posed problem in 3D reconstruction from the perspective of diffusion models and explain the underlying rationale. Next, we identify the challenges associated with using diffusion models to supplement scene-consistent details and our solutions. Finally, we present the intuition and analysis behind our solutions and their corresponding theoretical foundations. Our overall pipeline is shown in Figure~\ref{fig:pipeline}

% \subsection{Denoise Novel-view Rendering}
\subsection{Motivation and Challenge}
\label{sec:motivation&challenge}

%\KR{Change the assumption to approximate this error with Gaussian noise.}

Given the inherently ill-posed nature of 3D reconstruction tasks, rendering results from unseen viewpoints during training will inevitably deviate from the original scene details. We interpret this "deviation" as compensable by leveraging the prior knowledge embedded in diffusion models. Given a novel-view rendering $\mathbf{I}_n$ and its corresponding ground truth $\mathbf{I}_n^{g}$, their relationship can be written from an image-to-image denoising diffusion perspective, where 
\begin{equation}
\label{eq:noise_collection}
 \mathbf{I}_n^{g} = \mathbf{I}_n + \sum_{t \in T_{s}} { \bepsilon_\theta(\sqrt{\bar\alpha_t} \bx_t (\mathbf{I_n}, \bepsilon) + \sqrt{1-\bar\alpha_t}\bepsilon, t) }.
\end{equation}
Here, $T_s$ is a collection of reverse time steps constrained by broken strength $s_b$. Given the total inference time step $n$, $T_s=[t_1, t_2, ..., t_{n\times s_b}]$ includes only the first $n\times s_b$ steps. 

Despite the powerful generative capabilities of diffusion models, several challenges still arise in practical applications. (a) \textit{Selective Region Modification.} Not all regions require adjustment. The randomness of areas with deficiencies in novel viewpoints makes it difficult to precisely identify which regions need modification. Attempting to modify the
entire field of view can disrupt already well-reconstructed parts, leading to unintended distortions. (b) \textit{Multi-view Consistency.} The inherent randomness of noise in diffusion models often causes inconsistencies across different
viewpoints. This issue is particularly problematic in 3D reconstruction and has long been a major obstacle in previous research.

% \subsection{Further Analysis of Challenges}

\paragraph{Selective Region Modification.} Due to the explicit optimization strategy of 3D-GS, the parts that have already been well-trained but are modified can be easily refined in subsequent training iterations. In other words, as long as an appropriate training strategy is established, the regions appeared in the training dataset that can be fully reconstructed will remain largely unaffected. So we temporarily exclude this issue.

\paragraph{Multi-view Consistency.} We identify that the primary challenge affecting multi-view consistency lies in preserving finer details. Diffusion models could excel in recognizing and regenerating general scene and object content unbiasedly. However, the diversity introduced by noise in diffusion models presents significant difficulties in maintaining consistency at a finer detail level.

Acknowledging the inherent uncertainty introduced by diffusion models, we pivot to 3D-GS to investigate the manifestation of multi-view inconsistencies in this context, allowing us to refine the optimization target. Our investigation reveals that the \textbf{cloning process of Gaussian primitives} during optimization is the key to solving this issue. Our conclusion is as follows: by effectively controlling the gradients brought by the regeneration of diffusion models on Gaussian primitives across multiple viewpoints and regulating their cloning process, it is possible to produce detailed, consistent content across varying view angles. Detailed analysis is exhibited in the Appendix~\ref{sec:multi-view consistency}.

\subsection{Bootstrap Design}
% \label{sec:view_creation}
% \subsection{Bootstrap Design}
\label{sec:boot_design}
In our context, Bootstrap refers to utilizing partially reconstructed 3D scenes to extract novel-view renderings for regeneration and integration by diffusion models, leveraging
the existing content to enable self-improvement rather than relying on text-guided prompt generation. The overall pipeline is shown in Figure~\ref{fig:pipeline}.

%Our bootstrap pipeline begins by utilizing a trained 3D-GS model to sample novel-view renderings. These rendered images are then fed into the diffusion model for regeneration, serving as pseudo groundtruth to guide subsequent training. During training, for each training viewpoint, we strategically insert multiple bootstrapped viewpoints in a phased manner. This approach minimizes distortion in well-trained regions while promoting multi-view consistency in areas with deficiencies. 

\paragraph{Overall Diffusion Variance Control.} To ensure consistency across the overall scene, it is essential to initiate bootstrapping from a relatively complete stage of the reconstructed 3D scene, and only employ a small diffusion broken strength. This helps maintain the regenerated content without large variations.

On one hand, a well-trained 3D-GS model ensures that regions with deficiencies remain sufficiently close to the true scene, preventing significant biases in the diffusion process. Additionally, as outlined in~\cite{sohl2015deep, ho2020denoising}, the scale factor $\sqrt{1-\bar\alpha_t}$ in Equation~\ref{eq:noise_collection}, which determines the magnitude of added noise, approaches zero as the time step $t$ decreases to zero. By fixing the inference time step, larger time step values can be effectively excluded by constraining the breakage strength $s_r$, thereby preserving the regenerated scenes with minimal alteration.

\paragraph{Bootstrap Pipeline.} The core of our pipeline is to use multiple bootstrapped renderings in the same region to do average sampling to control the cloning gradient. We begin our bootstrapping by generating a set of new camera parameters for novel-view rendering (detailed in Appendix~\ref{sec:novel-view_creation}), where we make slight adjustments to the rotation and translation matrices of training cameras. Given a training image $I_t$, and its surrounding $k$ bootstrapped renderings $[I_t^{b_1}, ..., I_t^{b_k}]$, we then regenerate these renderings using a diffusion model and get $[I_t^{r_1}, ..., I_t^{r_k}]$. We use $\mathcal{L}_1$ loss to penalize the difference between $I_t^b$ and $I_t^r$, where the basic bootstrapping loss $\mathcal{L}_{b}= \lVert I_t^r - I_t^b \lVert$. During training, we incorporate all these bootstrapped renderings with the original training camera to make a hybrid loss
\begin{equation}
     \mathcal{L} = (1 - \lambda_{\text{boot}})\mathcal{L}_o + \frac {\lambda_{\text{boot}}}{k}  \sum_{i \in l} {\mathcal{L}^i_b},
     \label{eq:bootstrapping_loss}
\end{equation}
where $\mathcal{L}_o$ is the original 3D-GS training loss~\cite{kerbl20233dgs}. In practice, we bootstrap 2 variants for each training camera. And during training, we use the bootstrapped views not only from the current training camera but also the surrounding training cameras. This design choice is motivated by the common characteristic of most 3D reconstruction datasets~\cite{barron2022mipnerf360,hedman2018deep}, where cameras positioned in close proximity typically capture adjacent scenes. For example, if the training camera has both sides of surrounding training cameras, then the $l$ is 6.

\begin{figure}
    \centering
    \includegraphics[width=0.9\linewidth,height=5cm]{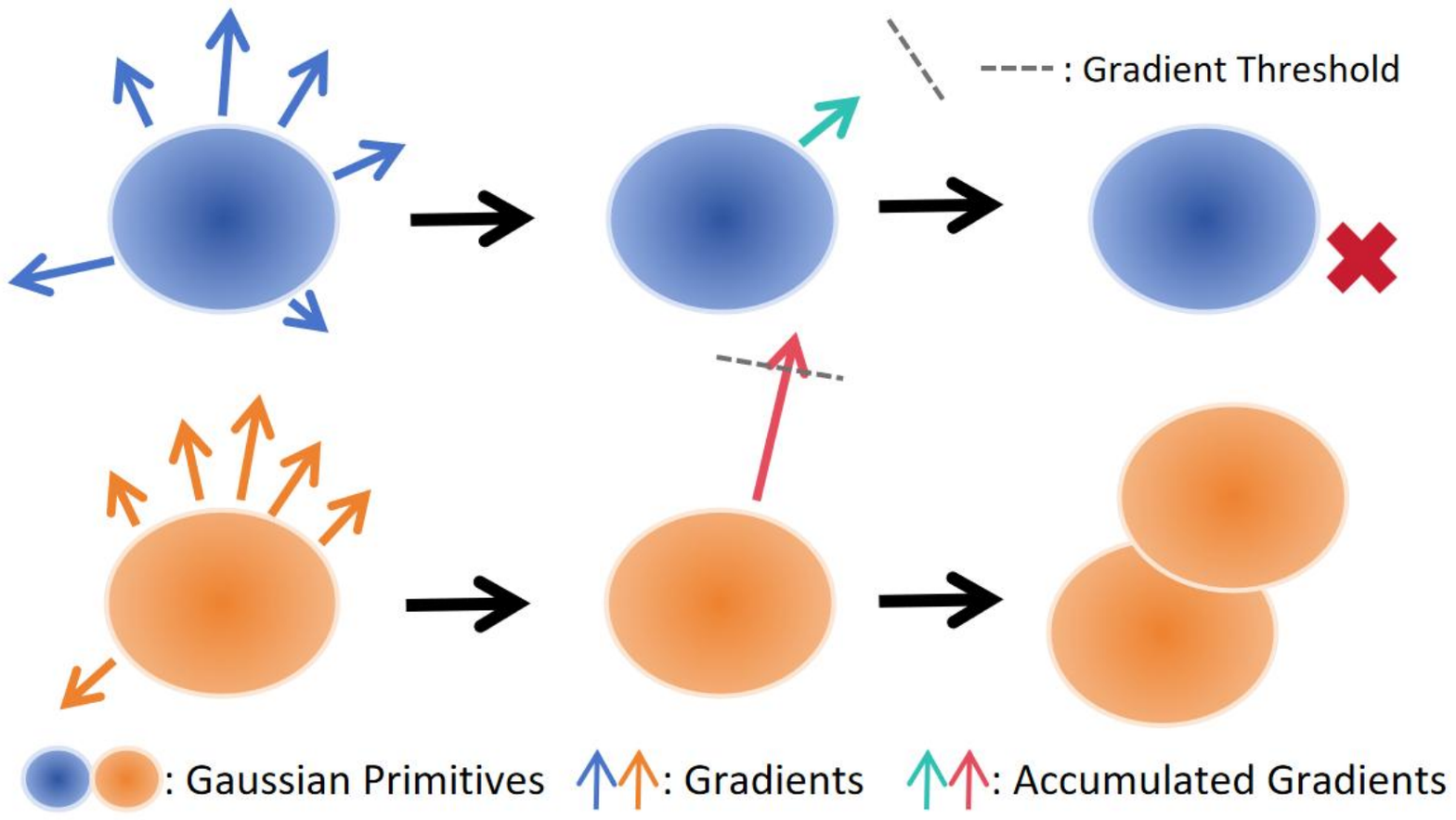}
    \caption{\textbf{3D-GS cloning process.} Only the gradients of Gaussian primitives aligned in nearly one direction have the potential to exceed the gradient threshold required for triggering further cloning.}
    \label{fig:gradient_accum}
    \vspace{-4pt}
\end{figure}

\paragraph{Diffusion Finetuning Strategies.}
In earlier analyses, we assumed that the diffusion model possesses robust generative capabilities. However, in practical applications, fine-tuning is often necessary to effectively denoise novel-view renderings, as the artifacts encountered in 3D reconstruction—such as distortion and aliasing—do not always conform to the noise distribution typically seen during diffusion model training. To address this, we adopt an \textbf{image-to-image} fine-tuning approach.

Our fine-tuning images are derived from trained 3D-GS models. Specifically, we use half of the training set to train 3D-GS and then utilize the model at the $4,000^{th}$–$6,000^{th}$ iterations (depending on the dataset) to render the remaining half of the training set. These renderings are identified as "broken" due to potential discrepancies and artifacts, with the corresponding aligned ground truths sourced from the other half of the training set. This process is repeated twice for each half of the training set, allowing us to generate hundreds of fine-tuning images across the entire dataset. Further details can be found in Appendix Sec.~\ref{sec:finetune}.

\subsection{Consistency Analysis}
\label{sec:consistency}

\paragraph{Diffusion Sampling Consistency.}
Ideally, a well-trained diffusion model with nonequilibrium-thermodynamics~\cite{sohl2015deep}, can reconstruct the original image from random noise through an infinite sequence of reverse diffusion sampling. As we have assumed the diffusion model meets the required performance, any observed variations can be attributed to the limitations of the finite time step schedule. Consider multi-sampling bootstrap loss term $\sum_{i \in N} {\mathcal{L}^i_b}$ in Equation~\ref{eq:bootstrapping_loss}, and combining it with Equation~\ref{eq:noise_collection}, we can reformulate it as
\begin{footnotesize}
\begin{align}
\label{eq:boot_loss_refor}
 \sum_{i \in k} {\mathcal{L}^i_b} \propto \sum_{i \in k} {\sum_{t \in T_{s_{r}}} { \bepsilon_\theta(\sqrt{\bar\alpha_t} \bx_t (\mathbf{I}_t^{b_i}, \bepsilon) + \sqrt{1-\bar\alpha_t}\bepsilon, t) }},
\end{align}
\end{footnotesize}
Given our focus on the imperfect segments, and considering that each segment's surrounding context is similar and homogeneous as described in Sec~\ref{sec:boot_design}, we can approximate each $\mathbf{I}_b^i$ within that segment as identical. Then, for each defective segment, we can rewrite Equation~\ref{eq:boot_loss_refor} as:
\begin{footnotesize}
\begin{align}
\label{eq:boot_loss_refor_1}
 \sum_{i \in k} {\mathcal{L}^i_b} \propto k\sum_{t \in T_{s_{r}}} { \bepsilon_\theta(\sqrt{\bar\alpha_t} \bx_t (\mathbf{I}_b, \bepsilon) + \sqrt{1-\bar\alpha_t}\bepsilon, t) },
\end{align}
\end{footnotesize}
Thus, by performing $k$-times repeated sampling, it mitigates the limitations arising from insufficient reverse diffusion sampling, ensuring a more robust reconstruction.

\paragraph{Noise Sampling Consistency.}
Since the diffusion model is a learned Gaussian noise predictor, we can also interpret the renderings from novel views as the results of adding noise with Gaussian distribution to the ground truths from a noise perspective. Then, we can rewrite $\sum_{i \in N} {\mathcal{L}^i_b}$ as a collection of noise samples under the same distribution (contextual similarity around the degraded parts) in accordance with Equation~\ref{eq:boot_loss_refor_1}:
\begin{footnotesize}
\begin{align}
\label{eq:boot_loss_noise}
 \sum_{i \in k} {\mathcal{L}^i_b} \propto \sum_{i \in k}{\epsilon_{I_t}^{b_i}} \approx k{\epsilon_{I_b}},
\end{align}
\end{footnotesize}
where $\sum_{i \in k}{\epsilon_{I_t}^{b_i}}$ and  $\epsilon_{I_b}$ are sampled from the same Gaussian distribution $\bepsilon_{I_b} \sim \mathcal{N}(\mu_{I_b}, \sigma_{I_b})$, corresponding to an imperfect segment. Under our assumption, since all $\epsilon_{I_t}^{b_i}$ represent the same region, their distributions are identical. Then utilizing Chebyshev's inequality to interpret Equation~\ref{eq:boot_loss_noise}, it becomes evident that this multiple noise sampling approach drives the expectation $\mathbb{E}[\sum_{i \in k}{\epsilon_{I_t}^{b_i}}]$ toward the mean $\mu_{I_b}$, while still preserving flexibility for controlled, diverse generation.

\paragraph{Practical Multi-view Consistency.} Multi-view consistency requires that the newly generated Gaussians from the clone process align with the existing scene, where gradients play a crucial role. From 3D-GS~\cite{kerbl20233dgs}, the training parameters of 3D-GS are optimized simultaneously yet separately, allowing us to consider a generalized situation.

Experimentally, the diffusion model typically generates images that are contextually aligned but differ in minor details. By setting an appropriate threshold for triggering further cloning, as illustrated in Figure~\ref{fig:gradient_accum}, a more stable optimization outcome can be achieved. Only gradients generally oriented in a single direction have the potential to exceed this threshold, resulting in a relatively faithful and context-aligned bootstrapping cloning point. In practice, instead of directly modifying the cloning threshold, we constrain the scaling factor $\lambda_{\text{boot}}$ to limit the bootstrapping gradient in Equation~\ref{eq:bootstrapping_loss}, which equals to control the cloning threshold while being more flexible to adjust.

\begin{table*}[htbp]
\centering
\renewcommand{\arraystretch}{1.15}
\setlength{\tabcolsep}{1pt}
\caption{\textbf{Quantitative comparison on real-world datasets.} Our bootstrapping compressed the number and total volume of Gaussian primitives and still significantly improved performance metrics. We have highlighted the \textbf{best} and \underline{second-best} results in each category.}
\vspace{-6pt}
\label{tab:real_q}
\resizebox{1\linewidth}{!}{
\begin{tabular}{l|cccc|cccc|cccc}
\toprule
Dataset & \multicolumn{4}{c|}{Mip-NeRF360} & \multicolumn{4}{c|}{Tanks\&Temples} & \multicolumn{4}{c}{Deep Blending} \\
\hline
\begin{tabular}{c|c} Method & Metrics \end{tabular}  & PSNR\(\uparrow\) & SSIM\(\uparrow\) & LPIPS\(\downarrow\) & \#GS(k)/Mem & PSNR\(\uparrow\) & SSIM\(\uparrow\) & LPIPS\(\downarrow\) & \#GS(k)/Mem & PSNR\(\uparrow\) & SSIM\(\uparrow\) & LPIPS\(\downarrow\) & \#GS(k)/Mem \\
\midrule

Mip-NeRF360~\cite{barron2022mipnerf360} & 27.69 & 0.792 & 0.237 & - & 23.14 & 0.841 & 0.183 & - & 29.40 & 0.901 & 0.245 & - \\

Mip-Splatting~\cite{yu2023mipsplat} & 27.61 & 0.816 & \textbf{0.215} & 1013/838.4M & 23.96 & 0.856 & 0.171 & 832/500.4M & 29.56 & 0.901 & 0.243 & 410/736.8M \\

\midrule
2D-GS~\cite{huang20242dgs} & 26.80 & 0.794 & 0.260 & \underline{391}/439.6M & 23.16 & 0.828 & 0.214 & \underline{361}/197.6M & 29.43 & 0.899 & 0.259 & \textbf{204}/349.4M \\

Our-2D-GS & 27.24 &0.804 & 0.255& \textbf{369}/390.3M & 23.51 & 0.839 & 0.209 & \textbf{324}/179.0M & \underline{31.18} & 0.915 & 0.247 & \underline{184}/314.5M \\

\midrule
3D-GS~\cite{kerbl20233dgs} & 27.55 & 0.813 & 0.220 & 799/629.3M & 23.70 & 0.852 & 0.169 & 711/371.5M & 29.76 & 0.907 & \underline{0.239} & 336/585.3M \\

Our-3D-GS & \underline{28.01} & \textbf{0.826} & \underline{0.218} & 671/522.9M & \underline{24.35} & \underline{0.858} & \underline{0.174} & 565/292.6M & 31.12 & \underline{0.916} & \textbf{0.233} & 263/451.0M \\

\midrule

Scaffold-GS~\cite{lu2023scaffold} & 27.73 & 0.812 & 0.226 & 643/\underline{170.9M} & 24.18 & 0.853 & 0.175 & 389/\underline{53.9M} & 30.16 & 0.908 & 0.252 & 923/\underline{76.1M} \\

Our-Scaffold-GS & \textbf{28.17} & \underline{0.824} & 0.221 & 625/\textbf{150.3M} & \textbf{24.78} & \underline{0.861} & \textbf{0.173} & 381/\textbf{53.8}M & \textbf{31.35} & \textbf{0.917} & 0.245 & 831/\textbf{50.3M} \\

\bottomrule
\end{tabular}}
\vspace{-5pt}
\end{table*}

\begin{figure*}[!t]
    \centering
    \includegraphics[width=1\linewidth]{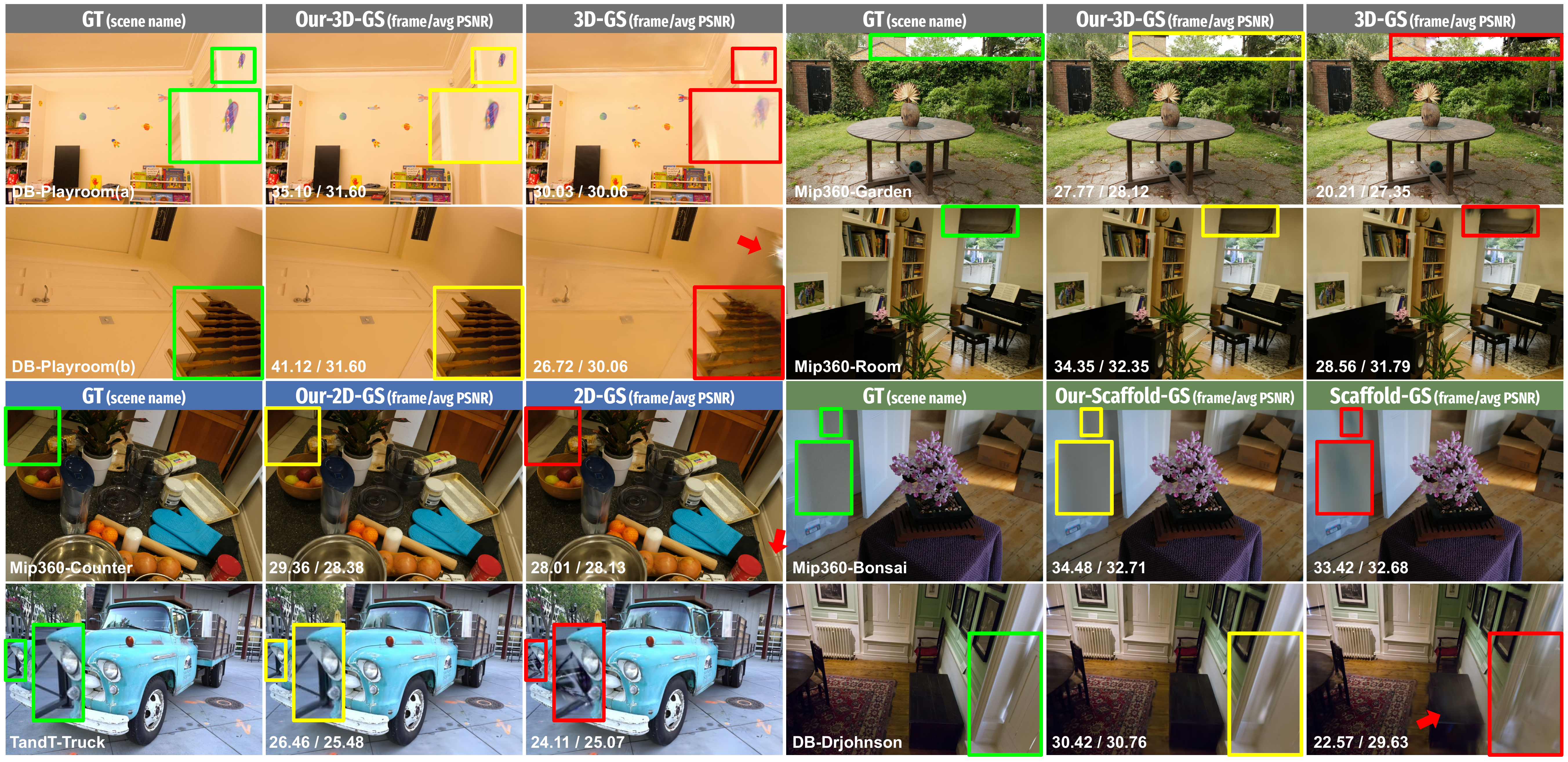}
    \caption{\textbf{Main comparisons.} Our bootstrapping pipeline successfully assisted the original baseline in denoising, enhancing details, filling in gaps, restoring distortions, and eliminating high-noise Gaussian primitives in novel views.}
    \vspace{-5pt}
    \label{fig:real_world_comp}
\end{figure*}

\vspace{-0.2cm}
\section{Experiment}

\subsection{Experimental Setup}
\paragraph{Dataset and Metrics.}
Our experiments are conducted on publicly available benchmark 3D reconstruction datasets in a verity of scenarios. They include 9 scenes from Mip-NeRF360~\cite{barron2022mipnerf360}, two scenes from Tanks\&Temples~\cite{knapitsch2017tanks}, two scenes from DeepBlending~\cite{hedman2018deep}, 8 scenes from BungeeNeRF~\cite{xiangli2022bungeenerf}, and 2 scenes from VR-NeRF~\cite{xu2023vr} (EyefulTower). For evaluation, we report PSNR, SSIM~\cite{wang2004image}, LPIPS~\cite{zhang2018unreasonable}, the number of used Gaussian primitives during rendering (\#GS), and the memory of models. We present the averaged results for all scenes in the main paper, while details are provided in the Appendix.

\paragraph{Baseline.}
We compare our method against the original 3D-GS~\cite{kerbl20233dgs}, Mip-Splatting~\cite{yu2023mipsplat}, Scaffold-GS~\cite{lu2023scaffold}, and 2D-GS~\cite{huang20242dgs}. We also report the results of MipNeRF360~\cite{barron2022mipnerf360} for rendering quality comparisons. To ensure fair comparisons with the original results, we kept all modeling parameters unchanged and only additionally incorporated our plug-and-play bootstrapping pipeline.

\paragraph{Implementation Details.}
\label{sec:imple_detail}
The novel-view bootstrapping begins at the $6k^{th}$ iteration for 3D-GS and Scaffold-GS, and at the $12k^{th}$ iteration for 2DGS. It is performed every 3k iterations until the $27k^{th}$ iteration. Within each bootstrapping interval, only 1k iterations are dedicated to bootstrap, while the remaining 2k iterations are reserved for standard training. This approach not only allows the model to recover from distortions introduced by bootstrap in well-trained regions but also avoids misalignment between previously regenerated renderings and the currently updated renderings. 

The loss scaling term $\lambda_{\text{boot}}$ is set to $[0.25, 0.1]$ for all baselines, with $0.25$ applied during the first 500 bootstrapping iterations and $0.1$ for the final 500 bootstrapping iterations. The broken strength decreases linearly from $0.1$ to $0.01$ throughout training. The diffusion model we use is the open-source \textbf{SDXL-Turbo}~\cite{rombach2022high} after finetuning, as delineated in Sec.~\ref{sec:consistency}. The other configurations and additional explanations for these configurations are provided in the Appendix~\ref{sec:config_explanation}.

\subsection{Performance Analysis}

\paragraph{Quality Comparisons}
As a plug-and-play technique, when combined with the bootstrap pipeline, rendering results of different baselines all experienced substantial enhancement, as shown in Figure~\ref{fig:real_world_comp},~\ref{fig:teaser},~\ref{fig:eyeful}. By providing additional training views through the bootstrap pipeline, our technique successfully enhances details, deblurs images, corrects distortions, and reduces high-frequency noise in the original model. On the other hand, the performance metrics of different baselines have also significantly improved across various scenarios, as shown in Tables~\ref{tab:real_q},~\ref{tab:bungee},~\ref{tab:eyeful}. Notably, the improvements in PSNR and SSIM are particularly significant. The average PSNR increase across all baselines exceeds 0.4, while SSIM improves by more than 1\%. For 2D-GS, LPIPS also decreases by more than 0.01.

\begin{table}[]
\centering
\caption{\textbf{Time comparison.} We report our 30k training time consumption on \textbf{Nvidia-H800} with and without our bootstrapping technique on \textbf{Mip-NeRF360}~\cite{barron2022mipnerf360} dataset.}
%\label{tab:lod}
\vspace{-6pt}
\footnotesize
\renewcommand{\arraystretch}{1.1}
\setlength{\tabcolsep}{7pt}
\resizebox{\linewidth}{!}{
\begin{tabular}{l|ccc}
\toprule
Method & GS & Diffusion & Total \\
\hline
2D-GS~\cite{huang20242dgs} & 26 min & - & 26 min \\
Our-2D-GS & 29 min & 18 min & 47 min \\

\midrule
3D-GS~\cite{kerbl20233dgs} & 21 min & - & 21 min \\
Our-3D-GS & 23 min & 18 min & 41 min \\

\midrule
Scaffold-GS~\cite{lu2023scaffold} & 19 min & - & 19 min \\
Our-Scaffold-GS & 21 min & 18 min & 39 min \\
\bottomrule
\end{tabular}
\label{tab:time_comparison}
}
\vspace{-5pt}
\end{table}

\begin{figure}
    \centering
     \includegraphics[width=0.9\linewidth,height=5cm]{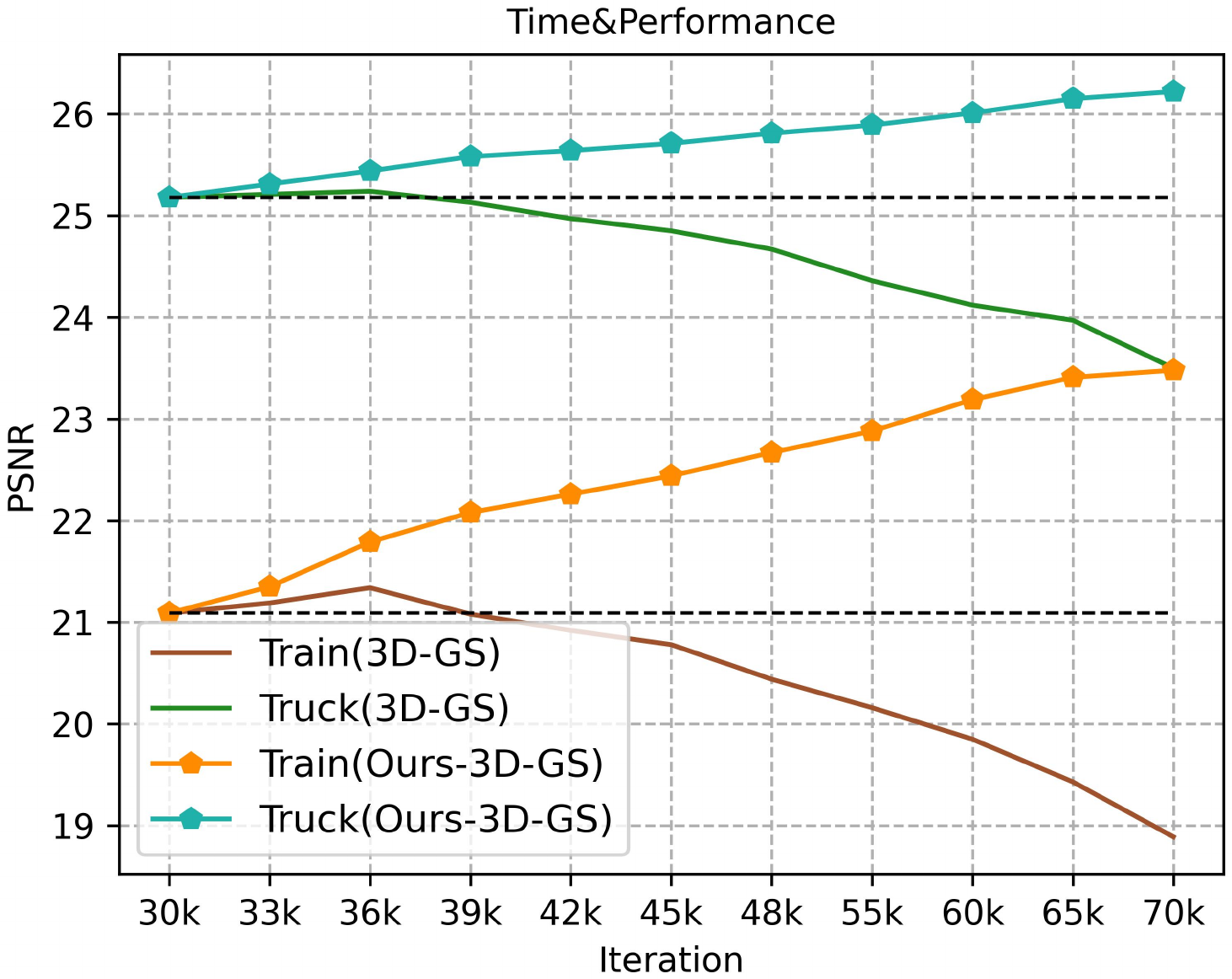}
    \vspace{-6pt}
    \caption{\textbf{Performance comparison on Tanks\&Temples datasets with extended training time.} While 3D-GS struggles to make further progress and even experiences a decline in performance, bootstrapping consistently enhances performance. The iterations are measured relative to the training process of 3D-GS.}
    \label{fig:time&perf}
    \vspace{-5pt}
\end{figure}

\paragraph{Storage Comparisons}
From Tables~\ref{tab:real_q}, \ref{tab:bungee}-\ref{tab:ablation_study}, we can observe that our bootstrapping can compress the model volume and reduce redundant Gaussians during rendering, especially for 2D-GS and 3D-GS. Compared to the original baselines, we achieved superior results using over 10\% fewer Gaussian primitives and reducing the volume by more than 20\%. 

This capability primarily stems from our multi-view sampling bootstrap loss in Equation~\ref{eq:bootstrapping_loss}. During training, cloning is based on the average gradient generated by Gaussian primitives under each view, where the bootstrapped views are also included in the total rendering count of these points. Since the averaged bootstrap scaler $\frac {\lambda_{\text{boot}}}{k}$ is very small in our setting, the introduction of new bootstrap views significantly suppresses the \textit{averaged gradients} produced by the original training loss under each view, thereby reducing the final volume and redundant points.

%\paragraph{Variants Comparisons}
%The gains brought by the bootstrap pipeline are effective across all baselines used in our experiments. Unlike traditional approaches that modify model structures or rendering methods, we enhance reconstruction results by addressing the scarcity of training data samples, making our method universally applicable to 3D reconstruction problems. Whether for the original 3D-GS or its variants, our technique has demonstrated significant improvements.

\begin{table}[tb]
    \centering
    \footnotesize
    \renewcommand{\arraystretch}{1.15}
    \setlength{\tabcolsep}{3pt}
    \caption{\textbf{Results on BungeeNeRF dataset.} Our bootstrapping pipeline helps other baselines achieve better performances while maintaining fewer rendering Gaussian primitives and volume.}
    \resizebox{1\linewidth}{!}{
    \begin{tabular}{l|cccc}%p{2cm}p{1cm}p{1cm}p{1cm}
    \toprule
    \begin{tabular}{c|c} Method & Metrics \end{tabular}  & PSNR\(\uparrow\) & SSIM\(\uparrow\) & LPIPS\(\downarrow\) & \#GS(k)/Mem \\
    \hline

    Mip-Splatting~\cite{yu2023mipsplat} & 28.14 & 0.918 & \textbf{0.094} & 2502/1610.2M\\

    \midrule

    2D-GS~\cite{huang20242dgs} & 27.08 & 0.900 & 0.124 & \underline{1099}/822.4M \\
    
    Our-2D-GS & 27.30 & 0.905 & 0.126 & \textbf{970}/708.2M \\
    
    \midrule
    
    3D-GS~\cite{kerbl20233dgs} & 27.68 & 0.915 & 0.098	& 2571/1656.4M \\
    
    Our-3D-GS & \textbf{28.37} & \textbf{0.924} & \underline{0.095} & 2017/1160.1M \\
    
    \midrule
    
    Scaffold-GS~\cite{lu2023scaffold} & 27.97 & 0.915 & 0.104 & 1202/\underline{168.0M} \\
    
    Our-Scaffold-GS & \underline{28.33} & \underline{0.920} & 0.101 & 1209/\textbf{165.8M} \\
    \bottomrule
    \end{tabular}}
    \vspace{-8pt}
    \label{tab:bungee}
\end{table}

\subsection{Efficiency Analysis}
\paragraph{Training Time Comparisons}
Since bootstrapping involves regenerating a large number of views using the diffusion model and rendering numerous novel perspectives, it inevitably increases training time to some extent. We conduct our time comparisons on an \textbf{Nvidia H800} device. On one hand, for standard training with 3k iterations, our training time nearly doubles, with the diffusion process accounting for the majority of the additional time, as shown in Table~\ref{tab:time_comparison}. However, the additional rendering overhead remains minimal due to the fast rasterization of 3D-GS and the exclusive use of \(\mathcal{L}_1\) loss for bootstrapping loss.

On the other hand, we also compare performance under the same total training time. In our setting, each round of bootstrapping is roughly equivalent to an additional 3k training iterations for a standard 3D-GS model. The results are shown in Figure~\ref{fig:time&perf}. After extended training time, the performance of 3D-GS \textbf{deteriorates rather than improves}. One fundamental reason is that model itself cannot compensate for the absence of new views. With prolonged training, while 3D-GS utilizes more Gaussian primitives to fit the training data, this excessive fitting instead leads to worse results for novel views. Only by filling in the missing new views, as our bootstrapping technique does, can the reconstruction results be further improved.

\subsection{Robustness Analysis}
\begin{figure}
    \centering
    \includegraphics[width=0.8\linewidth,height=5.5cm]{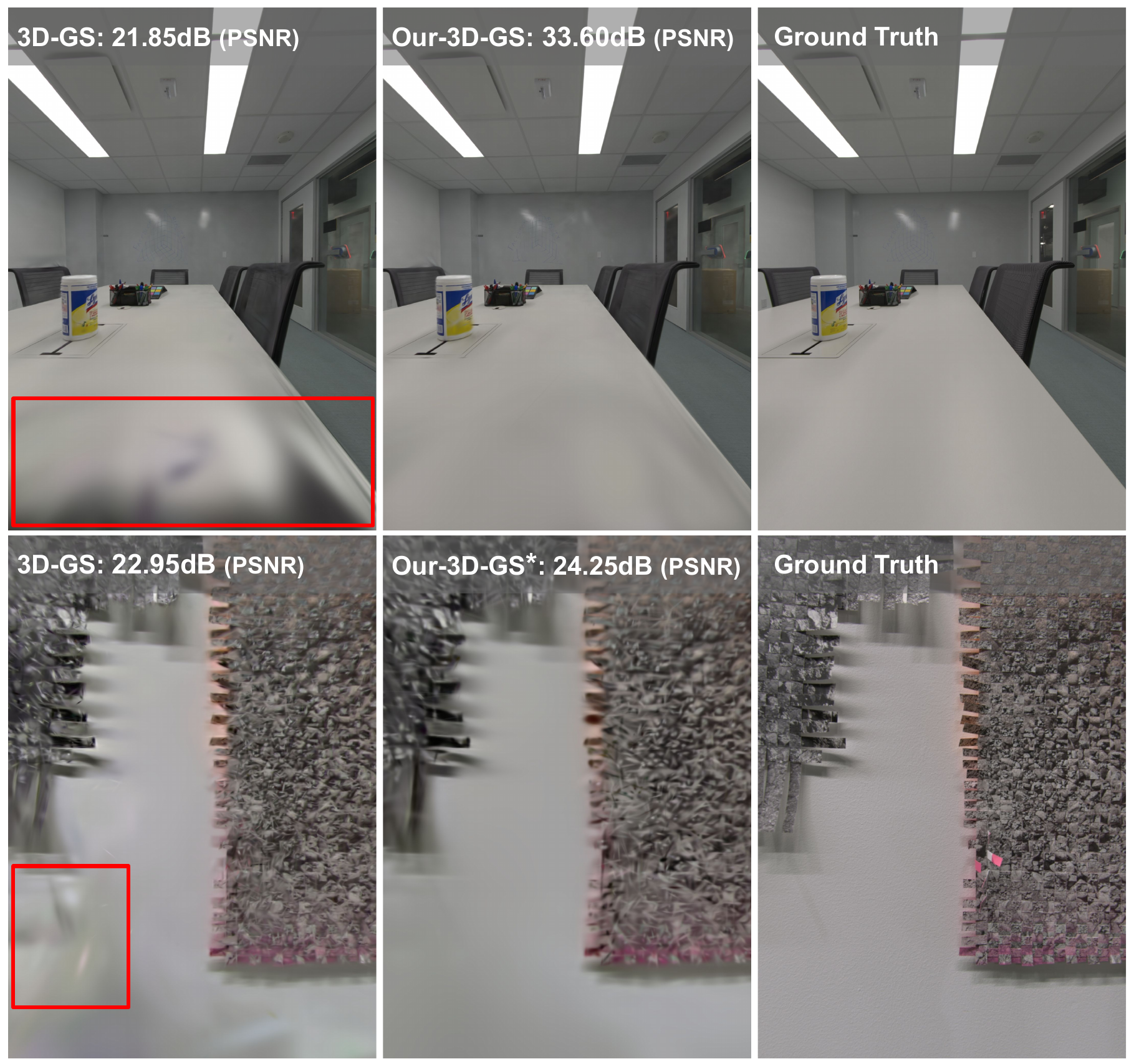}
    \vspace{-6pt}
    \caption{Comparisons on EyefulTower. * indicates we use unfine-tuned diffusion models in our bootstrap pipeline.}
    \label{fig:eyeful}
    \vspace{-8pt}
\end{figure}

\paragraph{Finetuning Results Analysis}
Our primary goal in fine-tuning is to better help the diffusion model fit the noise distribution under new problem settings. Through extensive experiments, we found that simply applying an unfine-tuned diffusion model to our bootstrap pipeline can still achieve scene denoising, content completion, and distortion correction in most scenarios, as shown in Figure~\ref{fig:eyeful} and Table~\ref{tab:eyeful}. This fully demonstrates the correctness of our theoretical analysis results and the stability of our method.

\paragraph{Multi-scale Results}
As shown in Table~\ref{tab:bungee}, our method maintains its superiority even when applied to the multi-scale dataset BungeeNeRF. This demonstrates that the multi-view consistency we emphasize throughout does not weaken due to the scale differences embedded in the datasets. For 3D-GS, our technique achieves a 0.7 PSNR improvement while using only 80\% of the original volume.

\begin{table}[tb]
    \centering
    %\setstretch{0.8}
    %\scriptsize
    \footnotesize
    \renewcommand{\arraystretch}{1.15}
    \setlength{\tabcolsep}{3pt}
    \caption{\textbf{Quantative performance on Eyefultower dataset.} * indicates we use unfine-tuned diffusion models in our bootstrap pipeline. Our pipeline delivers strong performance in large indoor scenes even without fine-tuning.}
    \vspace{-6pt}
    \resizebox{1\linewidth}{!}{
    \begin{tabular}{l|cccc}%p{2cm}p{1cm}p{1cm}p{1cm}
    \toprule
    \begin{tabular}{c|c} Method & Metrics \end{tabular}  & PSNR\(\uparrow\) & SSIM\(\uparrow\) & LPIPS\(\downarrow\) & \#GS(k)/Mem \\
    \hline
    
    3D-GS~\cite{kerbl20233dgs} & 31.36 & 0.927 &  0.239 & 363/463M \\

    Our-3D-GS* & 31.79 & 0.929 &  0.244 & 332/408M\\

    Our-3D-GS & 32.41 & 0.932 &  0.241 & 338/419M \\

    \bottomrule
    \end{tabular}}
    \vspace{-6pt}
    \label{tab:eyeful}
\end{table}

\paragraph{Large-scale indoor Results}
As shown in Table~\ref{tab:eyeful}, our method demonstrates high adaptability even in large-scale indoor datasets, where occlusions and complex lighting effects frequently occur. The improvements in rendering quality are evident, even without fine-tuning the diffusion model. As illustrated in Figure~\ref{fig:eyeful}, the bootstrapping pipeline effectively helps the model eliminate unwanted Gaussian primitives that contribute to high-frequency noise and distortions, leading to cleaner and more accurate reconstructions.

\subsection{Ablation Study}

Our ablation experiments focus on the proposed multi-view sampling loss in Equation~\ref{eq:bootstrapping_loss} and the finetuning method, as shown in Table~\ref{tab:ablation_study}. Here, \textbf{Finetune} refers to full-parameter finetuning using the traditional text2image approach, where the scene name serves as the keyword and the training set consists of corresponding images. \textbf{Unfinetuned} denotes the original \textbf{SDXL-Turbo}, while \textbf{LoRA} is derived from~\cite{hu2021lora}. When multi-view samples are not enough, the training results largely drop, together with the increment of Gaussian primitives and volume, confirming the validity of our theoretical reasoning and the effectiveness of our hybrid loss. Since the GS rasterization is fast enough, where additional n times' rendering onlyAdditionally, different finetuning methods also have significant impacts on the final results. Our proposed finetuning method outperforms both non-finetuned models and traditional finetuning methods, as this allows the diffusion model to better learn the distribution of noise corresponding to artifacts in 3D-GS reconstruction. It is worth noting that due to the overly complex training scenarios and insufficient training data, the results of full-parameter \textbf{Finetune} are even worse than \textbf{Unfinetuned} one.

\begin{table}[tb]
    \centering
    %\setstretch{0.8}
    %\scriptsize
    \footnotesize
    \renewcommand{\arraystretch}{1.15}
    \setlength{\tabcolsep}{5pt}
    \caption{\textbf{Ablation study on Mip-NeRF360 \textbf{Garden} of 3D-GS.} Boot-n refers to using $n$ surrounding bootstrap cameras to train a training camera during each training iteration.}
    \resizebox{1\linewidth}{!}{
    \begin{tabular}{l|cccc}%p{2cm}p{1cm}p{1cm}p{1cm}
    \toprule
    \begin{tabular}{c|c} Method & Metrics \end{tabular} & PSNR\(\uparrow\) & SSIM\(\uparrow\) & LPIPS\(\downarrow\) & \#GS(k)/Mem \\
    \hline
    Baseline & 22.54 & 0.635 & 0.346 & 701/711M \\
    \hline
    Bootstrap-1 view & 22.45 & 0.633 & 0.348 & 688/702M \\
    Bootstrap-2 views & 22.84 & 0.651 & 0.347 & 637/631M \\
    Bootstrap-4 views & 23.35 & 0.659 & 0.346 & 598/602M \\
    \midrule
    w/o Finetune & 22.67 & 0.639 & 0.348 & 624/653M \\
    T2I-Fintune & 22.31 & 0.629 & 0.350 & 643/685M \\
    LoRA-Fintune & 22.97 & 0.649 & 0.348 & 602/631M \\
    \midrule
    Ours (I2I-Fintune) & 23.48 & 0.665 & 0.345 & 571/569M \\
    \bottomrule
    \end{tabular}}
    \vspace{-7pt}
    \label{tab:ablation_study}
\end{table}

\section{Conclusion}
In this paper, we proposed a novel bootstrapping pipeline, which employs a diffusion model to compensate for the missing parts of the training scenario. The essence of our technique is its precise targeting and effective solutions to address the issue of training sampling deficiency in 3D reconstruction efforts. As a plug-and-play method, we demonstrated that with the integration of our pipeline, a variety of baselines achieve significant improvements in metrics. Additionally, our technique also refines the artifacts that are far divergent from training views, which are undetectable under normal views. 
\section*{Impact Statement}
Our work aims to address the long-standing ill-posed problem in 3D reconstruction—namely, the issue of missing training data samples. With the rapid advancements in diffusion models in recent years and the increasing number of works leveraging 2D prior knowledge to compensate for 3D gaps, it is natural to consider using diffusion models to aid in the reconstruction of real-world open scenes. However, before our work, no related study had attempted to directly integrate diffusion-generated content as training data into the 3D reconstruction process. \textbf{We are the first to accomplish this breakthrough.}

Our bootstrap pipeline is plug-and-play and applicable to any Gaussian Splatting-based approach, significantly enhancing the practical applicability of our work. Nearly all existing structural improvements to Gaussian Splatting frameworks are fully compatible with our data augmentation method. While previous works have achieved promising results on training and test datasets, they generally struggle with novel views that deviate significantly from training perspectives, highlighting the importance of our work for future Gaussian Splatting applications.

Moreover, the improvements shown in the images within our paper represent only a fraction of the enhancements our technique can provide. Since most novel views are not included in the test dataset, improvements in these unseen perspectives—such as the removal of noise clusters under a table in our pipeline~\ref{fig:pipeline}—are crucial for the practical application of Gaussian Splatting.

In summary, from the perspective of the importance of our work, the difficulty of the problem, and the innovation of our approach, we believe our work has achieved significant success in all these aspects. Furthermore, it provides invaluable support for the future industrial application of this field.
\bibliography{icml2025}
\bibliographystyle{icml2025}

%%%%%%%%%%%%%%%%%%%%%%%%%%%%%%%%%%%%%%%%%%%%%%%%%%%%%%%%%%%%%%%%%%%%%%%%%%%%%%%
%%%%%%%%%%%%%%%%%%%%%%%%%%%%%%%%%%%%%%%%%%%%%%%%%%%%%%%%%%%%%%%%%%%%%%%%%%%%%%%
% APPENDIX
%%%%%%%%%%%%%%%%%%%%%%%%%%%%%%%%%%%%%%%%%%%%%%%%%%%%%%%%%%%%%%%%%%%%%%%%%%%%%%%
%%%%%%%%%%%%%%%%%%%%%%%%%%%%%%%%%%%%%%%%%%%%%%%%%%%%%%%%%%%%%%%%%%%%%%%%%%%%%%%
\newpage
\appendix
%\onecolumn

\newpage
%\onecolumn
\section{Appendix}
\noindent \textbf{Overview.} This appendix is structured as follows:
(1) The first section elaborates on our further analyses and implementation details, (2) and additional experimental results are also presented.

%Given the inherently ill-posed nature of the 3D reconstruction tasks, it is inevitable that rendering results from unseen viewpoints during training will deviate from the original scene details. We interpret this "deviation" as a form of noise interference, assuming it follows a Gaussian distribution. Thus, for a rendered result $\mathbf{I}_n$ from a novel viewpoint and the corresponding ground-truth scene $\mathbf{I}_n^{g}$, we could express their relationship as 
%\begin{equation}
%\label{eq:noise_adding}
%    \mathbf{I}_n^{g} - \mathbf{I}_n \sim  \bepsilon_{I_n},
%\end{equation}
%where $\bepsilon_{I}^{n} \sim \mathcal{N}(\mu_{I_n}, \sigma_{I_n})$ represents the Gaussian noise. On the other hand, from Equation~\ref{eq:dm_obj}, we could rewrite the relationship between $\mathbf{I}_n^{g}$ and $\mathbf{I}_n$ from an image-to-image denoising diffusion perspective, where 

%If we adopt a single-step sampling approach using DDIM~\cite{song2020denoising} during diffusion, Equations~\ref{eq:noise_adding} and~\ref{eq:noise_collection} become essentially equivalent, where $\bepsilon_{I}^{n} = \bepsilon_\theta(\sqrt{\bar\alpha_t} \bx_t (\mathbf{I_n}, \bepsilon)+\sqrt{1-\bar\alpha_t}\bepsilon, t)$. This supports the use of diffusion models as a reasonable approach for compensating missing information in novel viewpoints.

\subsection{Emphasis on Cloning of Multi-view Consistency} 
\label{sec:multi-view consistency}
Our bootstrapping technique particularly emphasizes the cloning function in training. During the densification process, where points are either split or cloned, the surrounding points of these imperfect sections are initially flattened to compensate for the inconsistencies in the regenerated novel-view renderings. After several iterations' accumulation, big points are split into smaller ones, and small points are cloned along the gradient direction. In practical terms, these large points can be considered as coarse aggregations of smaller Gaussian points, where the renderings of these formations appear blurred. Over time, as the 3D-GS model progresses and refines, these big points gradually diminish and are refined into finer details. As such, they typically cease to exist in the final stages of the model's training. 

In conclusion, while large points may be noticeable in earlier phases, their overall impact on the final output of the model is minimal. Therefore, our primary focus should shift towards the cloning aspect. Specifically, we emphasize the gradient direction within the cloning procedure to ensure that finer details are accurately replicated and preserved.

When the $\lambda_\text{boot}$ is sufficiently small, it becomes evident that the bootstrapping loss term will have minimal effects on well-represented training cameras. For those underrepresented parts, the bootstrapping term can effectively facilitate modifications that remain consistent across multiple viewpoints. In our methodology, the bootstrapping scenarios consistently incorporate some of the same degenerated parts within our loss term configuration. Consequently, these parts are processed repeatedly from multiple angles, and the values are averaged over these areas to ensure uniformity and improve the overall quality of the synthesized views.

\subsection{Novel-view Creation}
\label{sec:novel-view_creation}
For each training camera, we construct 2 of its randomly generated cameras and put them back to training the same as in Bootstrap-GS. For random cameras, we altered both the rotation matrices $\bm{R}$ and the translation vectors $\bm{t}$ by adding random noise with scaling factors of 0.2 and 0.1, respectively (after which $\bm{R}$ was re-normalized to ensure it remained a valid rotation matrix).

\subsection{Configurations}
\label{sec:config_explanation}

\paragraph{Dataset Configurations}
We use \( llffhold = 8 \) for each dataset, meaning that for every span of 8 images, 1 image is designated for testing, while the remaining 7 are used for training.  

For most datasets, we conduct a total of 30k training iterations. However, for EyefulTower~\cite{xu2023vr}, due to its extensive 3D space, we extend the maximum training iterations to 60k. Correspondingly, the densification process is prolonged from $15k_{th}$ iteration to $30k_{th}$, while all other parameters of 3D-GS remain unchanged.

\paragraph{Further Explanations}
The decreasing setting of the diffusion broken strength $s_r$ is relatively simple to understand. As the training process advances, the model increasingly improves its representation of the reconstructed 3D scene, thus necessitating fewer alterations. For unfinetuned models, the maximum number of $s_r$ is set to $0.5$. But after finetuning, we suggest that changes made by the finetuned models should better align with the ground truth. Consequently, we can apply a greater $s_r$ in the earlier iterations to ensure that the initial corrections are more impactful.

In the two-stage approach to configuring $\lambda_\text{boot}$ outlined in Sec.~\ref{sec:imple_detail} of Experiments, our strategy is designed to address the deficiencies observed in 3D-GS. The original 3D-GS suffers from severe artifacts when rendering novel views under specific conditions. To mitigate this, we initially set a larger $\lambda_\text{boot}$ to introduce greater perturbations in the current renderings, enabling the generation of more context-aligned details for the diffusion model. Then, a small $\lambda_\text{boot}$ is applied to stabilize the outputs and ensure refinements for the subsequent generation of multi-view consistent details.

\subsection{Other Finetuning Strategies}
\label{sec:finetune}
Fine-tuning involves using the renderings as \( \mathbf{x}_t \) and the ground truths as \( \mathbf{x}_0 \). The process first introduces noise to the renderings to perturb their distribution, then performs noise sampling, and finally computes the loss against the ground truths. The sampling process is quite similar to the image-to-image generation of diffusion models~\cite{rombach2022high}.

Typically, it is recommended to use only a small fraction of the time steps, constrained by the broken strength \( s_b \), as defined in Sec.~\ref{sec:motivation&challenge}. This means limiting \( t \) in \( \mathbf{x}_t \) within the range \( T_s = [t_1, t_2, ..., t_{n \times s_b}] \). However, in practice, we adopt a higher broken strength during fine-tuning. For example, while \( s_b \) is set to $[0.1, 0.01]$ during bootstrapping, we increase it to $0.2$ or even $0.3$ during fine-tuning. This adjustment is particularly beneficial as distortions in the 3D-GS model can be significantly pronounced in certain areas, allowing the diffusion model to learn a more faithful and robust representation.

\subsection{Full Scene Results}

In this section, we present comprehensive details of our results on each scene, as shown in Tables 6-17 .

\begin{table}[htbp]
\renewcommand{\arraystretch}{1.1}
\setlength{\tabcolsep}{1pt}
\centering
\caption{PSNR for all scenes in the Mip-NeRF360\cite{barron2022mipnerf360} dataset.}
\vspace{-6pt}
\resizebox{\linewidth}{!}{
\begin{tabular}{l|ccccccccc}
\toprule
\begin{tabular}{c|c} Method & Scenes \end{tabular} & bicycle & bonsai &  counter & flowers & garden & kitchen & room & stump & treehill \\
\midrule

Mip-Splatting\cite{yu2023mipsplat} & 25.13 & 32.56 & 29.30 & 21.64 & 27.43 & 31.48 & 31.73 & 26.65 & 22.60 \\
\midrule
2D-GS~\cite{huang20242dgs} & 24.61 & 31.32 & 28.13 & 20.83 & 26.63 & 30.42 & 30.79 & 26.11 & 22.37 \\

Our-2D-GS & 24.95 & 31.43 & 28.38 & 21.20 & 27.24 & 30.62 & 31.47 & 26.78 & 23.09 \\
\midrule
3D-GS\cite{kerbl20233dgs} & 25.13 & 32.51 & 29.16 & 21.37 & 27.35 & 31.43 & 31.79 & 26.70 & 22.54 \\

Ours-3D-GS & 25.88 & 32.22 & 29.34 & 21.89 & 28.12 & 31.10 & 32.35 & 27.75 & 23.48 \\
\midrule
Scaffold-GS\cite{lu2023scaffold} & 25.00 & 32.68 & 29.66 & 21.31 & 27.29 & 31.80 & 32.16 & 26.63 & 23.02 \\

Ours-Scaffold-GS & 25.66 & 32.71 & 29.98 & 21.76 & 27.83 & 31.81 & 32.58 & 27.38 & 23.80 \\
\bottomrule

\end{tabular}
}
\end{table}
%--------------------------------------------
%--------------------------------------------

\vspace{-0.5em}
\begin{table}[htbp]
\renewcommand{\arraystretch}{1.1}
\setlength{\tabcolsep}{1pt}
\centering
\caption{SSIM for all scenes in the Mip-NeRF360\cite{barron2022mipnerf360} dataset.}
\vspace{-6pt}
\resizebox{\linewidth}{!}{
\begin{tabular}{l|ccccccccc}
\toprule
\begin{tabular}{c|c} Method & Scenes \end{tabular} & bicycle & bonsai &  counter & flowers & garden & kitchen & room & stump & treehill \\
\midrule
Mip-Splatting\cite{yu2023mipsplat} & 0.747 & 0.948 & 0.917 & \textbf{0.601} & 0.861 & 0.933 & 0.928 & 0.772 & 0.639 \\
\midrule
2D-GS\cite{huang20242dgs} & 0.713 & 0.935 & 0.899 & 0.556 & 0.834 & 0.921 & 0.914 & 0.752 & 0.619 \\

Ours-2D-GS & 0.724 & 0.937 & 0.902 & 0.572 & 0.845 & 0.924 & 0.920 & 0.773 & 0.640 \\
\midrule
3D-GS\cite{kerbl20233dgs} & 0.748 & 0.947 & 0.915 & 0.587 & 0.857 & 0.932 & 0.927 & 0.768 & 0.635 \\

Ours-3D-GS & 0.768 & 0.945 & 0.917 & 0.608 & 0.871 & 0.931 & 0.930 & 0.799 & 0.665 \\
\midrule
Scaffold-GS\cite{lu2023scaffold} & 0.742 & 0.948 & 0.918 & 0.579 & 0.850 & 0.932 & 0.930 & 0.763 & 0.642 \\

Ours-Scaffold-GS & 0.763 & 0.949 & 0.921 & 0.600 & 0.862 & 0.935 & 0.933 & 0.786 & 0.67 \\
\bottomrule

\end{tabular}
}
\end{table}
%--------------------------------------------
%--------------------------------------------

\vspace{-0.5em}
\begin{table}[htbp]
\renewcommand{\arraystretch}{1.1}
\setlength{\tabcolsep}{1pt}
\centering
\caption{LPIPS for all scenes in the Mip-NeRF360\cite{barron2022mipnerf360} dataset.}
\vspace{-6pt}
\resizebox{\linewidth}{!}{
\begin{tabular}{l|ccccccccc}
\toprule
\begin{tabular}{c|c} Method & Scenes \end{tabular} & bicycle & bonsai &  counter & flowers & garden & kitchen & room & stump & treehill \\
\midrule

Mip-Splatting\cite{barron2022mipnerf360} & 0.245 & 0.178 & 0.179 & 0.347 & 0.115 & 0.115 & 0.192 & 0.232 & 0.334 \\
\midrule
2D-GS\cite{huang20242dgs} & 0.306 & 0.205 & 0.215 & 0.403 & 0.162 & 0.137 & 0.222 & 0.290 & 0.398 \\

Ours-2D-GS & 0.302 & 0.203 & 0.213 & 0.396 & 0.159 & 0.136 & 0.217 & 0.280 & 0.391 \\
\midrule
3D-GS~\cite{kerbl20233dgs} & 0.241 & 0.179 & 0.182 & 0.359 & 0.121 & 0.116 & 0.195 & 0.242 & 0.346
\\

Our-3D-GS & 0.240 & 0.182 & 0.182 & 0.357 & 0.118 & 0.117 & 0.194 & 0.228 & 0.345
\\
\midrule
Scaffold-GS\cite{lu2023scaffold} & 0.259 & 0.179 & 0.181&0.370&0.134&0.117&0.188&0.260&0.342
\\

Ours-Scaffold-GS & 0.251& 0.176& 0.180& 0.365& 0.131& 0.117& 0.185& 0.250& 0.337 \\
\bottomrule

\end{tabular}
}
\end{table}
%--------------------------------------------
%--------------------------------------------

\vspace{-0.5em}
\begin{table}[htbp]
\renewcommand{\arraystretch}{1.1}
\setlength{\tabcolsep}{1pt}
\centering
\caption{Number of Gaussian Primitives(\#K) for all scenes in the Mip-NeRF360\cite{barron2022mipnerf360} dataset.}
\vspace{-6pt}

\resizebox{\linewidth}{!}{
\begin{tabular}{l|ccccccccc}
\toprule
\begin{tabular}{c|c} Method & Scenes \end{tabular} & bicycle & bonsai &  counter & flowers & garden & kitchen & room & stump & treehill \\
\midrule
Mip-Splatting\cite{yu2023mipsplat} & 1584 & 430 & 545 & 950 & 2089 & 1142 & 405 & 1077 & 892 \\
\midrule
2D-GS\cite{huang20242dgs} & 549& 220& 243& 374& 727& 432& 205& 394& 377 \\

Ours-2D-GS & 501& 212& 232& 359& 672& 426& 192& 387& 340 \\
\midrule
3D-GS\cite{kerbl20233dgs} & 1385& 360& 487& 725& 1471& 945& 317& 800& 701 \\

Ours-3D-GS & 1123& 323& 401& 587& 1246& 835& 273& 681& 571 \\
\midrule
Scaffold-GS~\cite{lu2023scaffold} & 765& 545& 383& 567& 1108& 851& 254& 631& 685 \\

Our-Scaffold-GS & 733& 517& 400& 568& 1105& 836& 248& 588& 630 \\
\bottomrule

\end{tabular}
}
\end{table}
%--------------------------------------------
%--------------------------------------------

\vspace{-2em}
\begin{table}[htbp]
\renewcommand{\arraystretch}{1.1}
\setlength{\tabcolsep}{1pt}
\centering
\caption{Storage memory(\#MB) for all scenes in the Mip-NeRF360\cite{barron2022mipnerf360} dataset.}
\vspace{-6pt}

\resizebox{\linewidth}{!}{
\begin{tabular}{l|ccccccccc}
\toprule
\begin{tabular}{c|c} Method & Scenes \end{tabular} & bicycle & bonsai &  counter & flowers & garden & kitchen & room & stump & treehill \\
\midrule
Mip-Splatting\cite{yu2023mipsplat} & 1433.6 & 318.1 & 307.5 & 970.2 & 1448.9 & 463.4 & 401.0 & 1239.0 & 964.3 \\
\midrule
2D-GS\cite{huang20242dgs} & 791.0& 186.4& 147.4& 492.2& 608.2& 191.8& 194.0& 685.7& 660.5\\

Our-2D-GS & 697.0& 178.0& 140.0& 462.0& 579.0& 182.0& 180.0& 591.0& 504.0 \\
\midrule
3D-GS\cite{kerbl20233dgs} 1176.0& 251.0& 247.07& 670.0& 999.0& 365.0& 296.0& 949.0& 711.0 \\

Our-3D-GS &932.0 & 216.0& 207.0& 541.0& 870.0& 323.0& 245.0& 803.0& 569.0 \\
\midrule
Scaffold-GS\cite{lu2023scaffold} & 277.0 & 123.0& 84.1& 208.0& 223.0& 101.0& 83.4& 226.0& 213.0 \\

Our-Scaffold-GS & 226.0& 111.0& 78.5& 191.0& 206.0& 96.3& 74.6& 187.0& 182.0 \\
\bottomrule

\end{tabular}
}
\end{table}
%--------------------------------------------
%--------------------------------------------

\vspace{-2em}
\begin{table}[htbp]
\centering
\renewcommand{\arraystretch}{1.05}
\setlength{\tabcolsep}{1pt}
\caption{Quantitative results for all scenes in the Tanks\&Temples\cite{knapitsch2017tanks} dataset.}
\vspace{-6pt}
\resizebox{\linewidth}{!}{
\begin{tabular}{l|cccc|cccc}
\toprule
Dataset & \multicolumn{4}{c|}{Truck} & \multicolumn{4}{c}{Train} \\

\begin{tabular}{c|c} Method & Metrics \end{tabular}  & PSNR  & SSIM  & LPIPS  & \#GS(k)/Mem & PSNR  & SSIM  & LPIPS  & \#GS(k)/Mem \\
\midrule
Mip-Splatting\cite{yu2023mipsplat} & 25.74 & 0.888 & 0.142 & 967/718.9M & 22.17 & 0.824 & 0.199 & 696/281.9M \\
\midrule
2D-GS\cite{huang20242dgs} & 25.07& 0.869& 0.175& 392/264.8M & 21.25& 0.788& 0.253& 331/130.4M \\

Our-2D-GS & 25.48& 0.878& 0.172& 355/242.1M & 21.54& 0.799& 0.247& 293/116.3M \\
\midrule
3D-GS\cite{kerbl20233dgs} & 25.52 & 0.885& 0.142& 751/486.0M & 21.87& 0.818& 0.196& 671/257.6M \\

Our-3D-GS & 25.83& 0.887& 0.150& 595/380.5M & 22.88& 0.829& 0.198& 534/205.3M \\
\midrule
Scaffold-GS\cite{lu2023scaffold} &25.86& 0.885& 0.143& 398/48.1M & 22.49& 0.821& 0.206& 380/59.6M \\

Our-Scaffold-GS &26.29& 0.892& 0.140& 396/48.6M & 23.27& 0.830& 0.205& 365/55.4M \\
\bottomrule

\end{tabular}
}
\end{table}

%--------------------------------------------
%--------------------------------------------

\vspace{-2em}
\begin{table}[htbp]
\centering
\renewcommand{\arraystretch}{1.05}
\setlength{\tabcolsep}{1pt}
\caption{Quantitative results for all scenes in the DeepBlending\cite{hedman2018deep} dataset.}
\vspace{-6pt}

\resizebox{\linewidth}{!}{
\begin{tabular}{l|cccc|cccc}
\toprule
Dataset & \multicolumn{4}{c|}{Dr Johnson} & \multicolumn{4}{c}{Playroom} \\

\begin{tabular}{c|c} Method & Metrics \end{tabular}  & PSNR  & SSIM  & LPIPS  & \#GS(k)/Mem & PSNR  & SSIM  & LPIPS  & \#GS(k)/Mem \\
\midrule
Mip-Splatting\cite{yu2023mipsplat} & 29.08 & 0.900 & 0.241 & 512/911.6M & 30.03 & 0.902 & 0.245 & 307/562.0M \\
\midrule
2D-GS\cite{huang20242dgs} & 28.89 & 0.898& 0.259& 246/415.6M & 29.97& 0.900& 0.261& 162/283.3M \\

Our-2D-GS & 30.47& 0.911& 0.246& 217/367.2M & 31.89& 0.919& 0.247& 152/262.8M \\
\midrule
3D-GS\cite{kerbl20233dgs} & 29.46& 0.905& 0.236& 424/735.2M & 30.06& 0.91& 0.241& 247/435.7M \\

Our-3D-GS & 30.64& 0.916& 0.228& 331/567.0M & 31.60& 0.915& 0.237& 195/335.0M \\
\midrule
Scaffold-GS\cite{lu2023scaffold} & 29.63& 0.905& 0.252& 936/59.2M & 30.70& 0.911& 0.251& 910/92.9M \\

Our-Scaffold-GS & 30.76& 0.913& 0.249& 967/53.8M & 31.93& 0.92& 0.241& 695/46.7M \\
\bottomrule

\end{tabular}
}
\end{table}
%--------------------------------------------
%--------------------------------------------

\vspace{-2em}
\begin{table}[htbp]
\renewcommand{\arraystretch}{1.1}
\setlength{\tabcolsep}{1pt}
\centering
\caption{PSNR for all scenes in the BungeeNeRF\cite{xiangli2022bungeenerf} dataset.}
\vspace{-6pt}

\resizebox{\linewidth}{!}{
\begin{tabular}{l|ccccccccc}
\toprule
\begin{tabular}{c|c} Method & Scenes \end{tabular}  & Amsterdam & Barcelona & Bilbao & Chicago & Hollywood & Pompidou & Quebec & Rome\\
\midrule
Mip-Splatting\cite{yu2023mipsplat} & 28.16 & 27.72 & 29.13 & 28.28 & 26.59 & 27.71 & 29.23 & 28.33 \\
\midrule
2D-GS\cite{huang20242dgs} & 27.22& 26.84& 28.57& 25.56& 26.31& 26.68& 28.40& 27.04 \\

Our-2D-GS & 27.31& 26.95& 28.75& 25.69& 26.91& 27.10& 28.45& 27.22 \\
\midrule
3D-GS\cite{kerbl20233dgs} & 27.67& 27.39& 28.79& 28.08& 26.19& 27.04& 28.74& 27.57 \\

Our-3D-GS & 28.27& 28.05& 29.37& 28.75& 27.35& 27.61& 29.12& 28.40 \\
\midrule
Scaffold-GS\cite{lu2023scaffold} & 28.11& 27.65& 29.25& 28.41& 26.39& 27.06& 28.88& 28.01 \\

Our-Scaffold-GS & 28.34& 27.91& 29.54& 28.63& 26.82& 27.69& 29.28& 28.40 \\
\bottomrule

\end{tabular}
}
\end{table}
%--------------------------------------------
%--------------------------------------------

\vspace{-2em}
\begin{table}[htbp]
\renewcommand{\arraystretch}{1.1}
\setlength{\tabcolsep}{1pt}
\centering
\caption{SSIM for all scenes in the BungeeNeRF\cite{xiangli2022bungeenerf} dataset.}
\vspace{-6pt}

\resizebox{\linewidth}{!}{
\begin{tabular}{l|ccccccccc}
\toprule
\begin{tabular}{c|c} Method & Scenes \end{tabular} & Amsterdam & Barcelona & Bilbao & Chicago & Hollywood & Pompidou & Quebec & Rome\\
\midrule
Mip-Splatting\cite{yu2023mipsplat} & 0.918 & 0.919 & 0.918 & 0.930 & 0.876 & 0.923 & 0.938 & 0.922 \\
\midrule
2D-GS\cite{huang20242dgs} & 0.893& 0.904& 0.910& 0.898& 0.869& 0.907& 0.923& \\

Our-2D-GS & 0.892& 0.906& 0.914& 0.901& 0.881& 0.916& 0.925& 0.903 \\
\midrule
3D-GS\cite{kerbl20233dgs} & 0.915& 0.917& 0.917& 0.930& 0.870& 0.918& 0.934& 0.917 \\

Our-3D-GS & 0.921& 0.924& 0.932& 0.936& 0.889& 0.925& 0.938& 0.926 \\
\midrule
Scaffold-GS\cite{lu2023scaffold} & 0.92& 0.916& 0.919& 0.927& 0.866& 0.917& 0.933& 0.919 \\

Our-Scaffold-GS & 0.922& 0.918& 0.923& 0.929& 0.879& 0.926& 0.936& 0.924 \\
\bottomrule

\end{tabular}
}
\end{table}
%--------------------------------------------
%--------------------------------------------

\vspace{-2em}
\begin{table}[htbp]
\renewcommand{\arraystretch}{1.1}
\setlength{\tabcolsep}{1pt}
\centering
\caption{LPIPS for all scenes in the BungeeNeRF\cite{xiangli2022bungeenerf} dataset.}
\vspace{-6pt}

\resizebox{\linewidth}{!}{
\begin{tabular}{l|ccccccccc}
\toprule
\begin{tabular}{c|c} Method & Scenes \end{tabular} & Amsterdam & Barcelona & Bilbao & Chicago & Hollywood & Pompidou & Quebec & Rome\\
\midrule
Mip-Splatting\cite{yu2023mipsplat} & 0.094 & 0.082 & 0.095 & 0.081 & 0.130 & 0.087 & 0.087 & 0.093 \\
\midrule
2D-GS\cite{huang20242dgs} & 0.136& 0.105& 0.114& 0.135& 0.155& 0.110& 0.114& 0.126 \\

Our-2D-GS & 0.143& 0.107& 0.114& 0.135& 0.152& 0.109& 0.115& 0.131 \\
\midrule
3D-GS\cite{kerbl20233dgs} & 0.099& 0.086& 0.097& 0.084& 0.134& 0.093& 0.092& 0.100 \\

Our-3D-GS & 0.099& 0.083& 0.096& 0.080& 0.125& 0.090& 0.091& 0.095 \\
\midrule
Scaffold-GS\cite{lu2023scaffold} & 0.097& 0.091& 0.100& 0.089& 0.162& 0.099& 0.095& 0.099 \\

Our-Scaffold-GS & 0.096& 0.092& 0.098& 0.082& 0.154& 0.096& 0.094& 0.095 \\
\bottomrule

\end{tabular}
}
\end{table}

%--------------------------------------------
%--------------------------------------------

\vspace{-0.5em}
\begin{table}[htbp]
\renewcommand{\arraystretch}{1.1}
\setlength{\tabcolsep}{1pt}
\centering
\caption{Number of Gaussian Primitives(\#K) for all scenes in the BungeeNeRF\cite{xiangli2022bungeenerf} dataset.}
\vspace{-6pt}

\resizebox{\linewidth}{!}{
\begin{tabular}{l|ccccccccc}
\toprule
\begin{tabular}{c|c} Method & Scenes \end{tabular} & Amsterdam & Barcelona & Bilbao & Chicago & Hollywood & Pompidou & Quebec & Rome\\
\midrule
Mip-Splatting\cite{yu2023mipsplat} & 2325 & 2874 & 2072 & 2712 & 2578 & 3233 & 1969 & 2251 \\
\midrule
2D-GS\cite{huang20242dgs} & 1040& 1264& 948& 1020& 1153& 1516& 876& 981 \\

Our-2D-GS & 851& 1132& 866& 936& 1073& 1305& 783& 814 \\
\midrule
3D-GS\cite{kerbl20233dgs} & 2311& 2970& 2057& 2648& 2752& 3474& 2070& 2290 \\

Our-3D-GS & 1804& 2317& 1623& 2076& 2224& 2584& 1647& 1862 \\
\midrule
Scaffold-GS\cite{lu2023scaffold} & 1527& 1206& 1099& 1360& 987& 1201& 945& 1294 \\

Our-Scaffold-GS & 1518& 1113& 1090& 1381& 1061& 1169& 1004& 1342 \\
\bottomrule

\end{tabular}
}
\end{table}

%--------------------------------------------
%--------------------------------------------

\vspace{-0.5em}
\begin{table}[htbp]
\renewcommand{\arraystretch}{1.1}
\setlength{\tabcolsep}{1pt}
\centering
\caption{Storage memory(\#MB) for all scenes in the BungeeNeRF\cite{xiangli2022bungeenerf} dataset.}
\vspace{-6pt}

\resizebox{\linewidth}{!}{
\begin{tabular}{l|ccccccccc}
\toprule
\begin{tabular}{c|c} Method & Scenes \end{tabular} & Amsterdam & Barcelona & Bilbao & Chicago & Hollywood & Pompidou & Quebec & Rome\\
\midrule
Mip-Splatting\cite{yu2023mipsplat} & 1464.3 & 1935.4 & 1341.4 & 1536.0 & 1607.7 & 2037.8 & 1382.4 & 1577.0 \\
\midrule
2D-GS\cite{huang20242dgs} & 699.2& 954.7& 682.5& 621.5& 783.3& 1310.4& 725.3& 802.9 \\

Our-2D-GS & 561.0& 799.0& 791.0& 555.0& 731.0& 992.0& 554.0& 683.0 \\
\midrule
3D-GS\cite{kerbl20233dgs} & 1470.5& 2029.6& 1317.1& 1507.3& 1697.3& 2173.9& 1439.6& 1616.1 \\

Our-3D-GS & 1094.0& 1442.0& 989.0& 1114.0& 1296.0& 1057.0& 1064.0& 1225.0 \\
\midrule
Scaffold-GS\cite{lu2023scaffold} & 205.6& 190.0& 161.9& 156.0& 139.6& 164.9& 145.3& 181.0 \\

Our-Scaffold-GS & 203.0& 185.0& 156.0& 151.0& 137.0& 177.0& 142.0& 175.0\\
\bottomrule

\end{tabular}
}
\end{table}
%%%%%%%%%%%%%%%%%%%%%%%%%%%%%%%%%%%%%%%%%%%%%%%%%%%%%%%%%%%%%%%%%%%%%%%%%%%%%%%
%%%%%%%%%%%%%%%%%%%%%%%%%%%%%%%%%%%%%%%%%%%%%%%%%%%%%%%%%%%%%%%%%%%%%%%%%%%%%%%

\end{document}